\documentclass[aip,amsmath,amssymb,reprint]{revtex4-1}

\usepackage{graphicx}
\usepackage{bm}

\usepackage{color}
\newcommand*{\revadd}[1]{{#1}}

\begin{document}

\title{Structural and dynamic properties of soda-lime-silica in the liquid phase}
\author{Alessandra Serva}
\affiliation{Sorbonne Universit\'e, CNRS, Physico-chimie des Electrolytes et Nanosyst\`emes Interfaciaux, PHENIX, F-75005 Paris, France}
\author{Allan Guerault}
\affiliation{Sorbonne Universit\'e, CNRS, Physico-chimie des Electrolytes et Nanosyst\`emes Interfaciaux, PHENIX, F-75005 Paris, France}
\affiliation{Surface du Verre et Interface (UMR 125), CNRS/Saint-Gobain Research Paris, 39 quai Lucien Lefranc, 93300 Aubervilliers, France}
\author{Yoshiki Ishii}
\affiliation{Graduate School of Simulation Studies,
University of Hyogo, 7-1-28 Minatojima-Minamimachi, Chuo-ku,
Kobe, Hyogo 650-0047, Japan}
\affiliation{Elements Strategy Initiative for Catalysts and 
Batteries,
Kyoto University, Katsura, Kyoto 615-8520, Japan}
\author{Emmanuelle Gouillart}
\affiliation{Surface du Verre et Interface (UMR 125), CNRS/Saint-Gobain Research Paris, 39 quai Lucien Lefranc, 93300 Aubervilliers, France}
\author{Ekaterina Burov}
\affiliation{Surface du Verre et Interface (UMR 125), CNRS/Saint-Gobain Research Paris, 39 quai Lucien Lefranc, 93300 Aubervilliers, France}
\author{Mathieu Salanne}
\email{mathieu.salanne@sorbonne-universite.fr}
\affiliation{Sorbonne Universit\'e, CNRS, Physico-chimie des Electrolytes et Nanosyst\`emes Interfaciaux, PHENIX, F-75005 Paris, France}
\affiliation{Institut Universitaire de France (IUF), 75231 Paris Cedex 05, France}

\begin{abstract}

Soda-lime-silica is a glassy system of strong industrial interest. In order to characterize its liquid state properties, we performed molecular dynamics simulations employing an aspherical ion model that includes atomic polarization and deformation effects. They allowed to study the structure and diffusion properties of the system at temperatures ranging from 1400 to 3000~K. We show that Na$^+$ and Ca$^{2+}$ ions adopt a different structural organization within the silica network, with Ca$^{2+}$ ions having a greater affinity for non-bridging oxygens than Na$^+$. We further link this structural behavior to their different diffusivities, suggesting that escaping from the first oxygen coordination shell is the limiting step for the diffusion. Na$^+$ diffuses faster than Ca$^{2+}$ because it is bonded to a smaller number of non-bridging oxygens. The formed ionic bonds are also less strong in the case of Na$^+$.

\end{abstract}

\maketitle

\section{Introduction}

A vast majority of manufactured glass are based on the soda-lime-silica system.
It is mainly made of three components: silica (SiO$_2$), which is a network
former setting the framework of the glass,~\cite{woodcock1976a} while calcium
oxide (CaO) and sodium oxide (Na$_2$O) are network modifiers~\cite{mead2006a,jund2001a} whose concentration strongly impacts many of the materials properties. A large number of additional cations, such as Al$^{3+}$, K$^+$, Mg$^{2+}$, Fe$^{3+}$, Ti$^{4+}$ are also present with very low concentrations. Due to the various uses of soda-lime-silica glass, such as windows and bottles to name the most common ones, its properties are very well characterized in the vitreous state.

The liquid state, which forms above $\approx$~1300~K, has far less been studied
due to the difficulty to conduct experiments at such temperatures. Indeed, the
literature on silicate melts at high temperature has focused more on
compositions of geological interest~\cite{zhang2007, giordano2008, zhang2010} than of industrial interest. A better characterization would be beneficial for many industrial processes. In
particular, understanding the impact of the sodium and the calcium ions on the
structure is of primary importance, and establishing the link between their
structure and their diffusion properties is necessary for a better control of
the glass melting and transformation properties. For example, it could provide
useful input for controlling  the rate of crystal
nucleation~\cite{fokin2006a,yuritsyn2020a} or the interdiffusion with
substrates~\cite{fonne2019a} during the glass formation. New experimental
setups have recently been proposed to study glassy oxides in the liquid
state,~\cite{ohkubo2012a,bolore2019a} but so far the only available
experimental data for diffusivities in liquid soda-lime-silica were obtained
through studies of trace elements using electrochemical
methods,~\cite{russel1991a} or isotopic tracer diffusivity in
undercooled melts, close to the glass transition.~\cite{njiokep2008}

In this respect, classical Molecular Dynamics (MD) simulations are a powerful
tool to deeply understand both structural and dynamics properties at a
molecular level. Many different force fields are dedicated to silica-derived
materials. The most generic ones employ fixed partial charges and a
Buckingham-type pair
potential,~\cite{pedone2006a,pedone2008a,laurent2014,molnar2013} but they are mostly used to study the structural and mechanical properties in the glassy state. For example, Guillot {\it et al.} have developed a series of potentials to study the thermodynamic properties of natural silicate melts (magmatic liquids) at very high temperature.~\cite{guillot2007a,guillot2007b,dufils2017a,dufils2018a} The soda-lime-silica structure was studied using such potentials, either from simulations only~\cite{cormack2001a} or through combination with diffraction experiments.~\cite{cormier2004a} Nevertheless, when studying the dynamics of such complex oxides, Madden and co-workers~\cite{madden1996a,jahn2006a,jahn2007b} have shown the importance of accounting for complex interactions, such as polarization effects as well as ionic deformation effects for the repulsion term, leading to the aspherical ion model (AIM). Such potentials are more computationally expensive, but they generally show a high level of agreement with experiments and a high transferability upon composition, temperature and pressure changes. The parameters can be obtained through force and dipole-fitting against first-principles calculation,~\cite{madden2006a} so that no experimental information is used during the development. In recent work, we have validated such a potential for sodium aluminosilicate glasses and melts against a large body of experimental data (bond lengths, neutron and X-ray diffraction, and NMR spectroscopy).~\cite{ishii2016a} The potential was then extended to MgO-SiO$_2$ and CaO-SiO$_2$ mixtures.~\cite{salmon2019a} In this study we use this AIM potential to investigate the structure and the dynamics of a model soda-lime-silica system in the liquid state.

\section{Model and methods}\label{methods}
\subsection{Aspherical ion model (AIM)}
The AIM \revadd{potential energy} consists in a sum of charge-charge, polarization, repulsion and dispersion terms.  
   \begin{eqnarray}
    \label{eq:pot}
    \phi^{\rm tot} =
    \phi^{\rm charge-charge} 
    + \phi^{\rm pol}
    + \phi^{\rm rep}
	   +\phi^{\rm disp}   
   \end{eqnarray}
\noindent All of them are provided (in atomic units) in the following. Firstly,
\begin{equation}
\phi^{\rm charge-charge}=\sum_i \sum_{j>i} \frac{q_{i}q_{j}} {r_{ij}}
\end{equation}
where $q_i$ is the charge of each ion $i$, and $r_{ij}$ the distance between $i$ and $j$. Contrarily to \revadd{many} rigid ion models,~\cite{pedone2006a,guillot2007a} formal charges are used for all the ions (-2 for O, +4 for Si, +3 for Al, +2 for Ca and +1 for Na in the present work). The polarization energy term ${\phi}^{\rm pol}$ includes charge-dipole and dipole-dipole interactions,
   \begin{eqnarray}
    \label{eq:pol}
    \phi^{\rm pol} = 
    \sum_i \sum_{j>i} \left[
	    \frac{ q_{i} {\bf r}_{ij} \cdot {\pmb \mu }_{j} } {r_{ij}^3} \revadd{f^{ij}_{D}}\left( r_{ij} \right)
	   - \frac{ {\pmb \mu }_{i} \cdot {\bf r}_{ij} q_{j} } {r_{ij}^3} \revadd{f^{ji}_{D}}\left( r_{ij} \right)
    \right] \\ \nonumber
    + \sum_i \sum_{j>i} \left[
    \frac{ {\pmb \mu }_{i} \cdot {\pmb \mu}_{j} } {r_{ij}^3}
    - \frac{ 3 ( {\bf r}_{ij} \cdot {\pmb \mu}_{i} ) 
             ( {\bf r}_{ij} \cdot {\pmb \mu}_{j} ) } {r_{ij}^5} \right]
    + \sum_i \frac{|{\pmb \mu}_i|^2}{2\alpha_i}
   \end{eqnarray}
where ${\pmb \mu}_i$ is the induced dipole moment of particle $i$.  The \revadd{$f_{D}^{ij}$} are functions~\cite{tang1984a} \revadd{that correct} the charge-dipole interactions \revadd{at short range},~\cite{wilson1996b,wilson1996c}
   \begin{equation}
	   \revadd{f^{ij}_D}(r_{ij}) = 1-\revadd{c_D^{ij}}{\rm e}^{-\revadd{b_D^{ij}} r_{ij}} 
	   \sum_{k=0}^4 \frac{(\revadd{b_D^{ij}} r_{ij})^k}{k!} \label{eq:tangtoennies1}
   \end{equation}
\revadd{in which $b_D^{ij}$ and $c_D^{ij}$ are two parameters giving the range and the strength of the correction. Note that the range of damping of the charge of $i$ on the dipole of $j$ is the same as the one of the charge of $j$ on the dipole of $i$ ($b_D^{ij}=b_D^{ji}$), but this reciprocity is not true for the correction strength ($c_D^{ij}\ne c_D^{ji}$). The effect of the latter on anions is generally to reduce the dipole from its asymptotic value~\cite{madden1996a,jemmer1999a} while the situation is more complex for cations dipoles.~\cite{domene2001a}} The induced dipoles are calculated by solving self-consistently the set of equations
\begin{equation}
{\pmb \mu }_{i}= \alpha_i {\bf E_i}\left(\{q_j\}_{j\ne i}, \{{\pmb \mu }_{j}\}_{j\ne i}\right),
\end{equation}
\noindent where ${\bf E_i}$ is the electric field generated at ${\bf r}_i$ by the whole set of charges and induced dipoles from the ions $j\ne i$, and $\alpha_i$ is the polarizability of ion $i$.  In practice, the instantaneous dipole moments are determined at each time step by minimization of the total energy \revadd{(at fixed positions and charges -- the latter being constant parameters throughout the whole simulation)} using the conjugate gradient method.~\cite{jahn2007a}  The charge-charge, charge-dipole and dipole-dipole contributions to the potential energy and forces of each ion are evaluated under the periodic boundary condition by using the Ewald summation technique.~\cite{aguado2003a}

The repulsive term is given by:~\cite{jahn2007b,salanne2012a}
\revadd{
   \begin{eqnarray}
    \label{eq:repaim}
	   \phi^{\rm rep} &=& 
    \sum_{i\in{\rm cation}} \sum_{j\in{\rm O}}
    \left[ A_1^{ij} \exp (-b_1^{ij} \rho_{ij}) +A_2^{ij} \exp (-b_2^{ij}\rho_{ij}) \right] \\ \nonumber
	   &+& \sum_{i\in{\rm cation}} \sum_{j\in{\rm O}} A_3^{ij} \exp (-b_3^{ij}r_{ij}) \\ \nonumber 
	   &+& \sum_{i\in{\rm O}} \sum_{j\in{\rm O}, i<j} A_1^{ij} \exp (-b_1^{ij} r_{ij}) \\ \nonumber
	   &+& \sum_{i\in{\rm O}}
    \left[ D \left\{\exp(\beta\delta\sigma_i)+\exp(-\beta\delta\sigma_i)\right\} \right. \\ \nonumber
	   &+& \left. \left\{ \exp(\zeta^2|\nu_i|^2)-1\right\}\right]
   \end{eqnarray}
   }
\noindent where $\rho_{ij}$ plays the role of an ``effective'' distance between ions $i$ and $j$. It is calculated at each time step using
   \begin{eqnarray}
    \rho_{ij}=r_{ij}-\delta\sigma_i-\delta\sigma_j-\frac{1}{r_{ij}}{\pmb r}_{ij}\cdot({\pmb \nu}_i-{\pmb \nu}_j),
   \end{eqnarray}
where $\delta\sigma_i$ represents the deviation of the ionic radius, and ${\pmb \nu}_i$ expresses the distortion of the dipolar shape.  These variables therefore account for the anisotropic deformation of the ionic shape. They are treated as additional degrees of freedom of the simulation. \revadd{As the induced dipoles, they are computed at each timestep by minimizing the total energy at fixed positions through a conjugate gradient procedure;}  the fourth summation term of eq~\ref{eq:repaim} consists of the self-energy terms to account for the energy cost of these deformation.  The repulsive interaction thus includes many-body effects. Finally, the dispersion term is given by an usual asymptotic expansion

\begin{equation}
	\phi^{\rm disp}=   - \sum_i \sum_{j>i} \left[ \frac{C^{ij}_6} {r_{ij}^6} f^{ij}_{6}\left( r_{ij} \right)+ \frac{C^{ij}_8} {r_{ij}^8} f^{ij}_{8}\left( r_{ij} \right) \right]
\end{equation}
\noindent where similar damping functions are used as for the charge-dipole term to screen the interactions at short range \revadd{, except that they involve only a single range parameter in each case:
   \begin{eqnarray}
	   f^{ij}_6(r_{ij}) = 1-{\rm e}^{-\revadd{b_6^{ij}} r_{ij}}\sum_{k=0}^6 \frac{(\revadd{b_6^{ij}} r_{ij})^k}{k!}\\
	   f^{ij}_8(r_{ij}) = 1-{\rm e}^{-\revadd{b_8^{ij}} r_{ij}}\sum_{k=0}^8 \frac{(\revadd{b_8^{ij}} r_{ij})^k}{k!}
   \end{eqnarray}
   }

\revadd{All the} parameters for such an AIM  \revadd{are fitted} from first-principles calculations \revadd{, except from the dispersion coefficients ($C_6^{ij}$ and $C_8^{ij}$) which were adjusted in order to reproduce experimental glass densities under ambient condition (because the functional used for the first-principles calculations notoriously underestimates dispersion effects~\cite{corradini2014a})}. \revadd{This parameterization was already performed} for Si, Al, Na and O in Ref. \citenum{ishii2016a} and for Ca in Ref. \citenum{salmon2019a}.  The \revadd{parameters} are listed in Tables~\ref{tb:pol1}, \ref{tb:pol} and~\ref{tb:rep}.

\begin{table}
  \caption{\label{tb:pol1}
	Individual parameters of the AIM force field. The Si and Al atoms are not polarizable \revadd{ in the AIM model (because their polarizabilities are negligible)}, and only the O atoms are deformable. All the parameters are provided in atomic units.}
    \begin{tabular} {cccccc} 
	\hline     
    \multicolumn{1}{c}{$i$} &
    \multicolumn{1}{c}{O}  &
    \multicolumn{1}{c}{Si} &
    \multicolumn{1}{c}{Al} &
    \multicolumn{1}{c}{Ca} &
    \multicolumn{1}{c}{Na} \\
	\hline     
    $q_{i}$    &     -2  &     +4  &   +3    &     +2  &     +1  \\
    $\alpha_{i}$& 10.74  &   --      &   --      & 3.183   &  0.991  \\
    \revadd{$D$}        & 0.5287  &   --      &    --      &   --      &   --      \\
    \revadd{$\zeta$}    & 1.6838  &   --      &    --      &   --      &   --      \\
    \revadd{$\beta$}    & 1.5723  &   --      &    --      &   --      &   --      \\
	\hline     
    \end{tabular}
\end{table}

\begin{table}
  \caption{\label{tb:pol}
	Pair parameters for charge-charge and polarization terms in soda-lime silicates (dipole damping functions \revadd{provided in Equation \ref{eq:tangtoennies1}}).  All the parameters are provided in atomic units.}
    \begin{tabular} {cccccc} 
	\hline     
    \multicolumn{1}{c}{$i$-$j$} &
    \multicolumn{1}{c}{O-O}  &
    \multicolumn{1}{c}{Si-O} &
    \multicolumn{1}{c}{Al-O} &
    \multicolumn{1}{c}{Ca-O} &
    \multicolumn{1}{c}{Na-O} \\
	\hline     
	    \revadd{$b_D^{ij}$}   &  2.513  &  1.939  &  1.908  &  1.769  &  1.964  \\
	    \revadd{$c_D^{ij}$}   &  2.227  &  1.446  &  1.627  &  1.881  &  \revadd{3.493}  \\
	    \revadd{$c_D^{ji}$}   &  2.227  &         &         &  0.144  &  \revadd{0.066}  \\
	\hline     
    \end{tabular}
\end{table}

\begin{table}
  \caption{\label{tb:rep}
	Pair parameters for repulsive and dispersion terms in soda-lime silicates.  All the parameters are provided in atomic units. The cations interact between themselves only through electrostatics (the corresponding repulsion terms are null).}
    \begin{tabular} {cccccc} 
	\hline     
    \multicolumn{1}{c}{$i$-$j$} &
    \multicolumn{1}{c}{O-O}  &
    \multicolumn{1}{c}{Si-O} &
    \multicolumn{1}{c}{Al-O} &
    \multicolumn{1}{c}{Ca-O} &
    \multicolumn{1}{c}{Na-O} \\
	\hline     
	\revadd{$A_1^{ij}$}   &  970.7  &  37.15  & 42.16    &  120.79 &  56.85  \\
	\revadd{$b_1^{ij}$}   &  2.674  &  1.499  & 1.592    &  1.763  &  1.738  \\
	\revadd{$A_2^{ij}$}   &         &  47863  & 13243    &  21256  &  23804  \\
	\revadd{$b_2^{ij}$}   &         &  9.921  & 4.400    &  9.020  &  4.081  \\
	\revadd{$A_3^{ij}$}   &         & 2930.9  & 2930.9   & 2930.9  & 2930.9  \\
	\revadd{$b_3^{ij}$}   &         &  3.906  & 3.906    &  3.906  &  3.906  \\
    $C^{ij}_6$ &   68.2  &    2.0  & 2.0      &    45.0  &   40.0  \\
    $C^{ij}_8$ &    783  &     25  &  25      &     450  &    400  \\
    $b^{ij}_6$ &    1.0  &    2.2  &  2.2     &    2.2  &    2.2  \\
    $b^{ij}_8$ &    1.0  &    2.2  &  2.2     &    2.2  &    2.2  \\
	\hline     
    \end{tabular}
\end{table}

\subsection{Molecular dynamics simulations}
Classical MD simulations of CaO-Na$_2$O-Al$_2$O$_3$-SiO$_2$ system with composition in mol\% 10.71 CaO, 14.53 Na$_2$O, 0.3 Al$_2$O$_3$ and 74.46 SiO$_2$ were carried out in a range of temperature 1400 K - 3000 K. The corresponding numbers of ions in the simulation box are 293 O$^{2-}$, 125 Si$^{4+}$, 1 Al$^{3+}$, 47 Na$^+$ and 18 Ca$^{2+}$.\\

The system was studied at six different temperatures (1400, 1800, 2000, 2200, 2400 and 3000~K). The simulation cells were first equilibrated by performing a simulation in the NPT ensemble\revadd{, using the method of Martyna {\it et al.}~\cite{martyna1994a} with relaxation times of 0.5~ps for both the thermostat and the barostat,} for at least 1~ns, with a time step of 0.5~fs. Production simulations were carried out in the NVT ensemble, using again a timestep of 0.5~fs and saving a configuration every 100 steps. The Nos\'{e}-Hoover chain thermostat,\cite{martyna1992a} with a relaxation constant of 0.5~ps, was used to control the system temperature. The total simulation time was 10~ns for the 1400~K, 1800~K, 2000~K and 2200~K temperatures and 6~ns for the 2400~K and 3000~K temperatures. \revadd{The short-ranged repulsion interactions were calculated with a cut-off distance fixed to half the simulation cell (typically 10~\AA) while the Ewald summation method was used for dispersion~\cite{karasawa1989a} and electrostatics~\cite{aguado2003a} interactions. It employed the same cut-off distance as for repulsion for the short-range sum, and an accuracy of 10$^{-7}$.}

\begin{figure*}[t!]
\begin{center}
\includegraphics[width=0.8\textwidth]{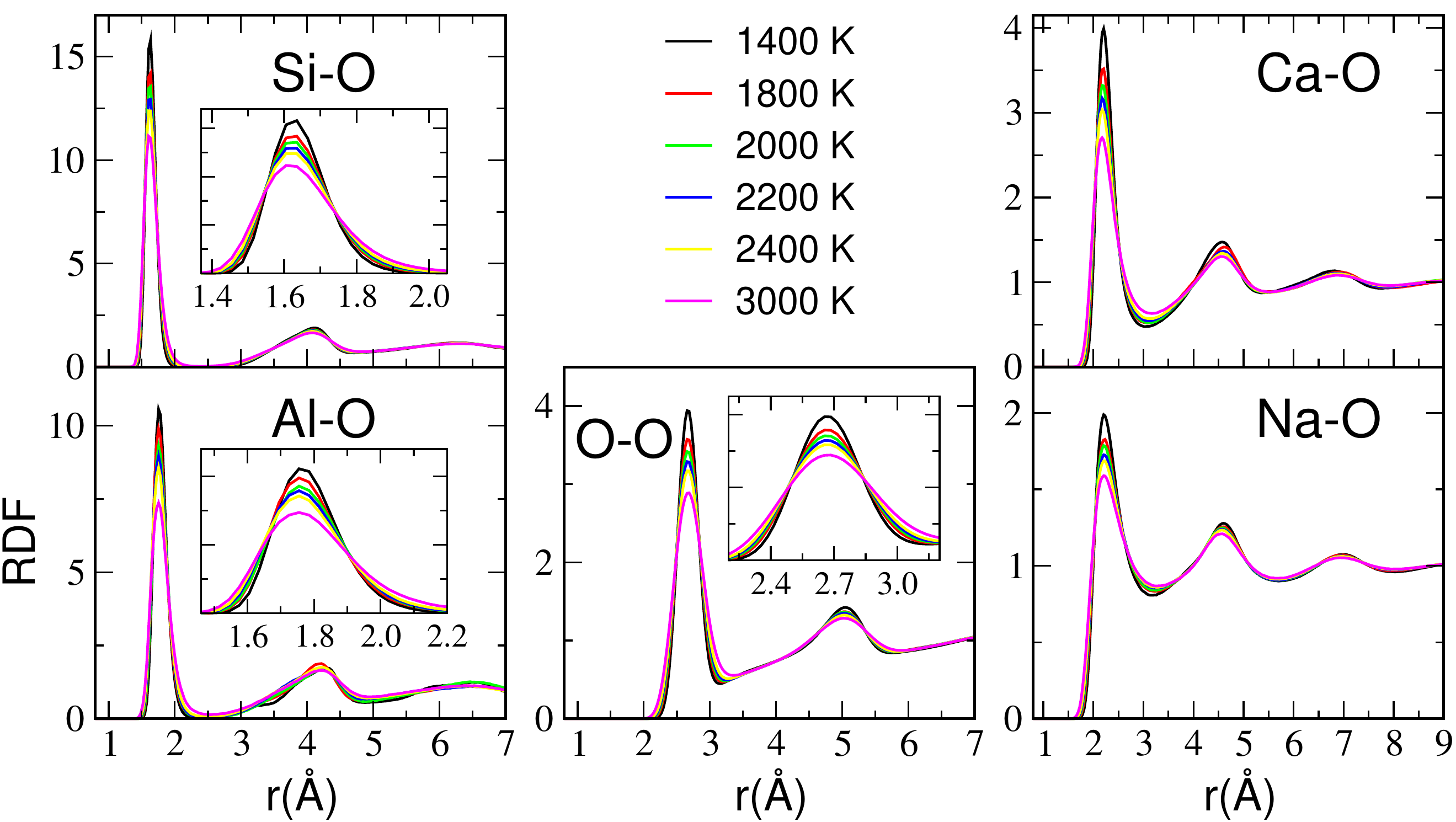}\\
\end{center}\caption{X-O RDFs calculated for the various temperatures (X = Si, Ca, Al, O or Na). The insets for the Si-O, Al-O and O-O pairs show a magnified view of the RDF first peaks.}\label{rdf1}
\end{figure*}

\section{Results and Discussions}\label{Result}

\subsection{Short-range structure}
We start the structural analysis by characterizing the coordination shell of the various cationic species. The radial distribution functions (RDFs) between the network formers Si and Al and the oxygen atoms (Si-O and Al-O) show very intense first peaks followed by a first minimum close to 0 (see Figure \ref{rdf1}, left panels). This is typical of a well-defined first coordination shell. The position of the first peak gives Si-O and Al-O distances 1.62 and 1.75 \AA, respectively. These distances do not vary with temperature, but there is a decrease in intensity and a broadening of the peak with increasing temperature. The corresponding average coordination number, calculated by integration of the RDF up to the first minimum, is 4 for both of them for all the investigated temperatures. These numbers are consistent with our previous simulation works on aluminosilicate systems,~\cite{ishii2016a} and experimental values of Ref.~\citenum{cormier2011a}.

\begin{figure*}[t]
\begin{center}
\includegraphics[width=0.8\textwidth]{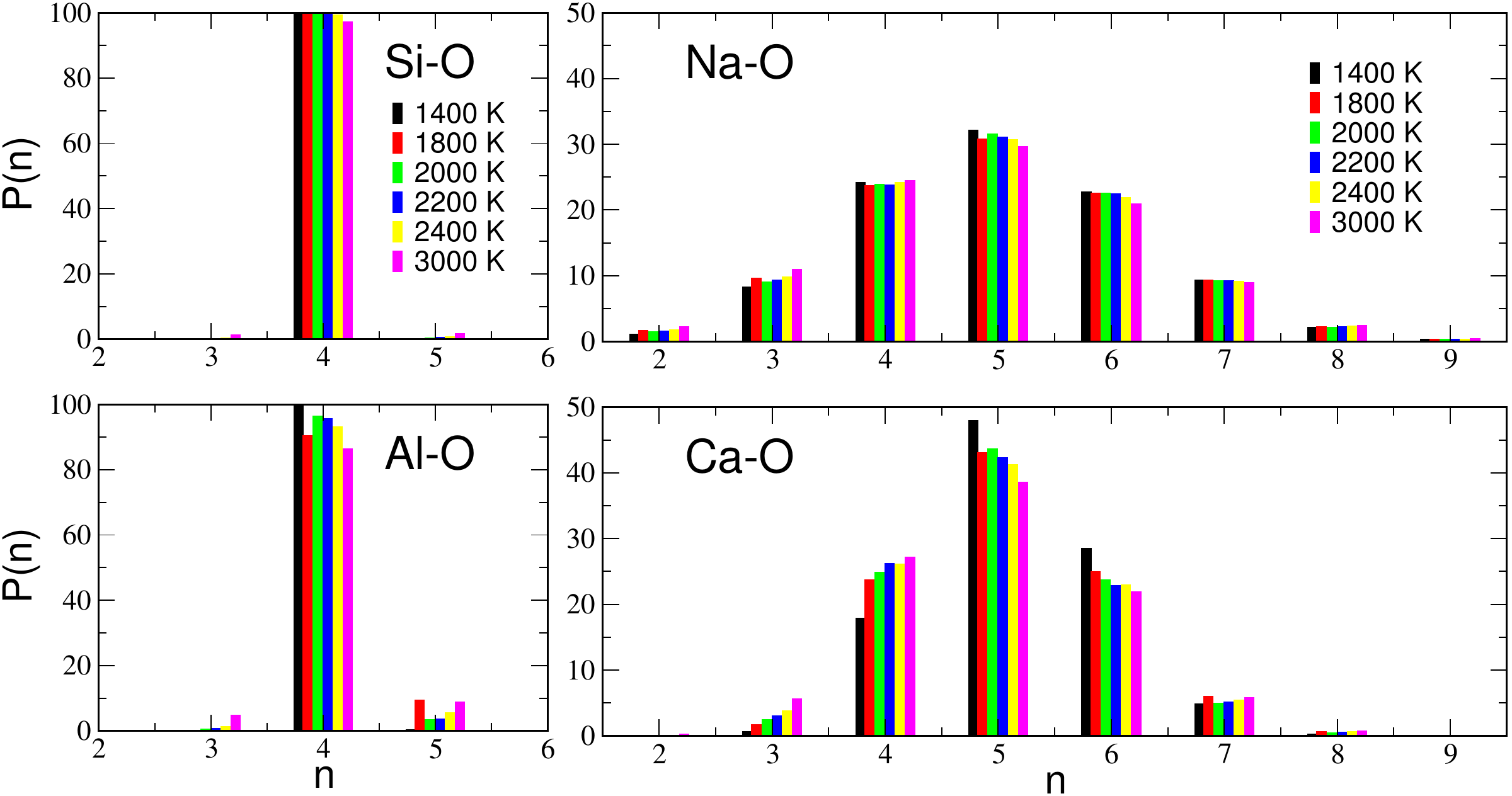}\\
\end{center}\caption{Instantaneous coordination number ($n$) probability distributions, expressed in percentage, for the various temperatures.}\label{ncoord}
\end{figure*}

\begin{figure}[t]
\begin{center}
\includegraphics[width=0.4\textwidth]{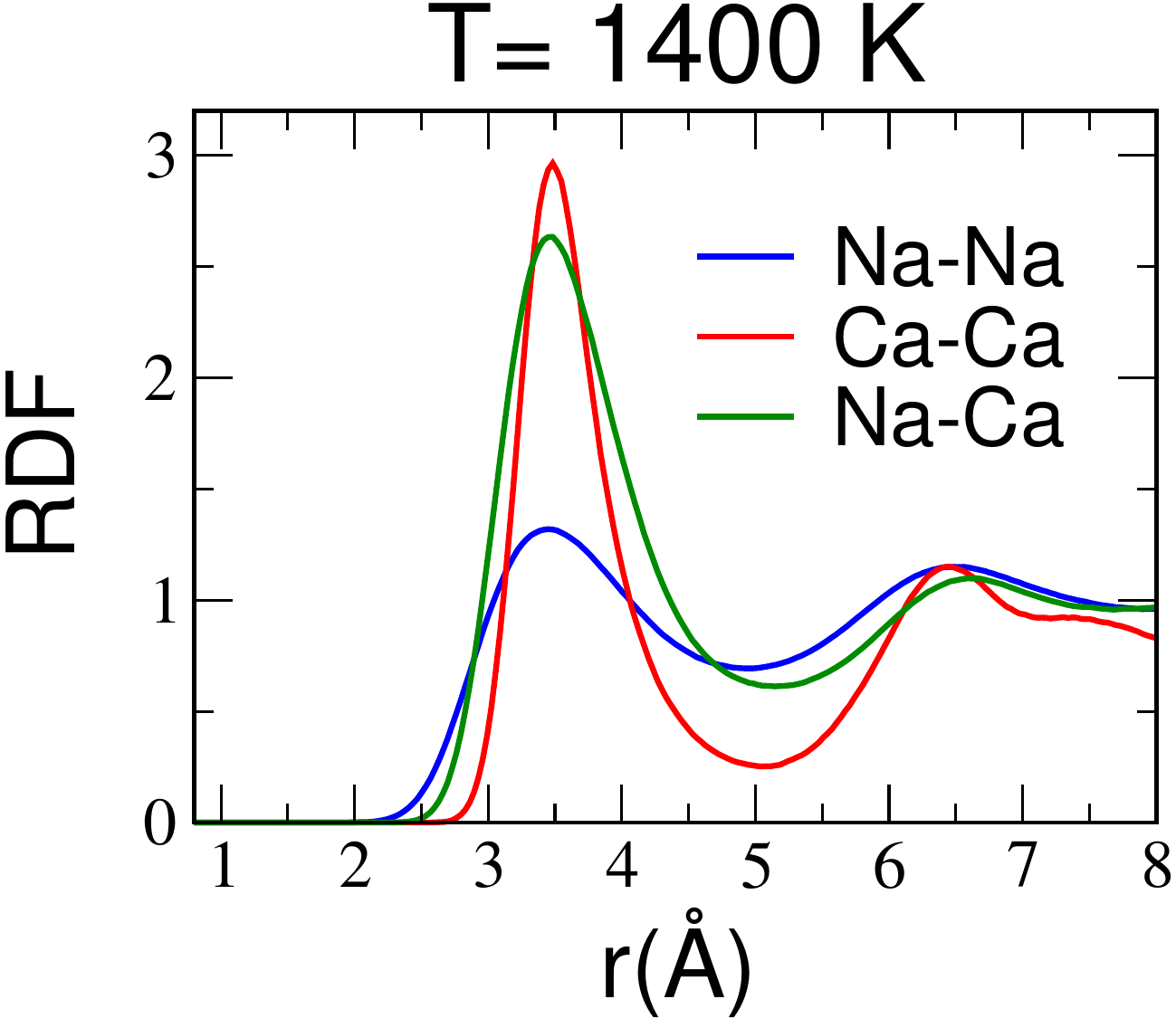}\\

\end{center}\caption{Na-Na, Ca-Ca and Na-Ca radial distribution functions (RDFs) calculated at 1400~K.}\label{gdr_1400}
\end{figure}

Figure \ref{rdf1} also shows the RDFs between the network modifiers, Na and Ca, and the oxygen atoms. We can see that the Na-O and Ca-O most likely distances are very close since the corresponding RDFs first peaks are respectively centered at 2.22 and 2.20~\AA, with an average coordination number of about 5 for both cations. These numbers are in agreement with previous theoretical studies of soda-lime-silicate system carried out at 300~K,\cite{cormack2001a,laurent2014} while a coordination number of about 7 for Ca and 6 for Na has been found in another XAS/MD study.\cite{cormier2004a} However, as discussed below, the distribution of coordination number is much wider than for Si and Al, which can explain such discrepancies. 
	The position of this first peak also does not change with increasing temperature. In particular, the Ca-O RDF's first peak is \revadd{somewhat} narrower as compared to the Na-O one, denoting a more rigid coordination shell for Ca$^{2+}$ ions. Anyway, in both cases the RDF's first minimum does not go to zero, meaning that the first coordination shell is more flexible in comparison to the network formers. \\
\indent This feature is reflected in the instantaneous coordination number analysis (see Figure \ref{ncoord}), which shows that the Si and Al cations are almost only 4-fold coordinated by oxygen atoms at all investigated temperatures, while Ca and Na cations can adopt a very large number of different coordinations. In particular, the Ca cations can take all values from 3 to 8, with 5 as favored configuration and 4/6 as second favored configurations. For Na cations the distribution of coordination numbers is even broader, ranging from 2 to 9, with preferential values of 4, 5 and 6. Moreover, the Na-O instantaneous coordination number probability seems not to be affected by the temperature, while a small trend can be seen for the Ca cations. For the latter the 5- and 6-fold coordinated species probabilities decrease with increasing temperature, while the 3- and 4-fold ones increase, with a consequent slight decrease of the average coordination number (from 5.2 at 1400 K to 5.0 at 3000 K).\\
\indent If we then look at long-ranged interactions, such as the ones occurring between Na-Na and Ca-Ca, the RDFs show a first contribution around 3.5~\AA. An example for the simulation at 1400~K is reported in Figure~\ref{gdr_1400}. According to the literature,\cite{cormack2001a} this distance is smaller than the one required for a homogeneous distribution and it is generally interpreted using the modified random network model with preferential regions concentrated with modifiers and non-bridging oxygens. Also the Na-Ca RDF shows a peak between 3 and 4~\AA, suggesting a mixing between Na and Ca atoms. In particular, the average coordination numbers are 2.50 for Na-Na, 1.48 for Na-Ca, 1.05 for Ca-Ca and 3.90 for Ca-Na. According to these numbers, 63\% (2.50/3.98*100) of the 3.98 network modifiers found on average around a Na atom are Na atoms while 37\% (1.48/3.98*100) are Ca atoms. In the same way, 21\% of the network modifiers around a Ca atom are Ca atoms while 79\% are Na atoms. If we assume a random distribution of Na and Ca in the glass and considering that at the investigated glass composition 72\% of the network modifiers are Na (N$_{Na}$/(N$_{Na}$+N$_{Ca}$)*100) and 28\% are Ca, we would expect to find around any network modifier 72\% of Na atoms and 28\% of Ca atoms. However, our numbers clearly show that around Na atoms there are less Na atoms (63\%) than expected from a random distribution (72\%), and  around Ca atoms there are less Ca atoms (21\%) than expected from a random distribution (28\%). Our result thus indicates a preference for Na and Ca to mix, in agreement with a previous experimental NMR study,\cite{lee2003a} where it has been shown a significant non random distribution of the Na and Ca modifying cations and this nonrandomness appears to be primarily governed by charge differences among cations. 

\subsection{Structure factors}

\begin{figure}[!h]
\begin{center}
\includegraphics[width=\columnwidth]{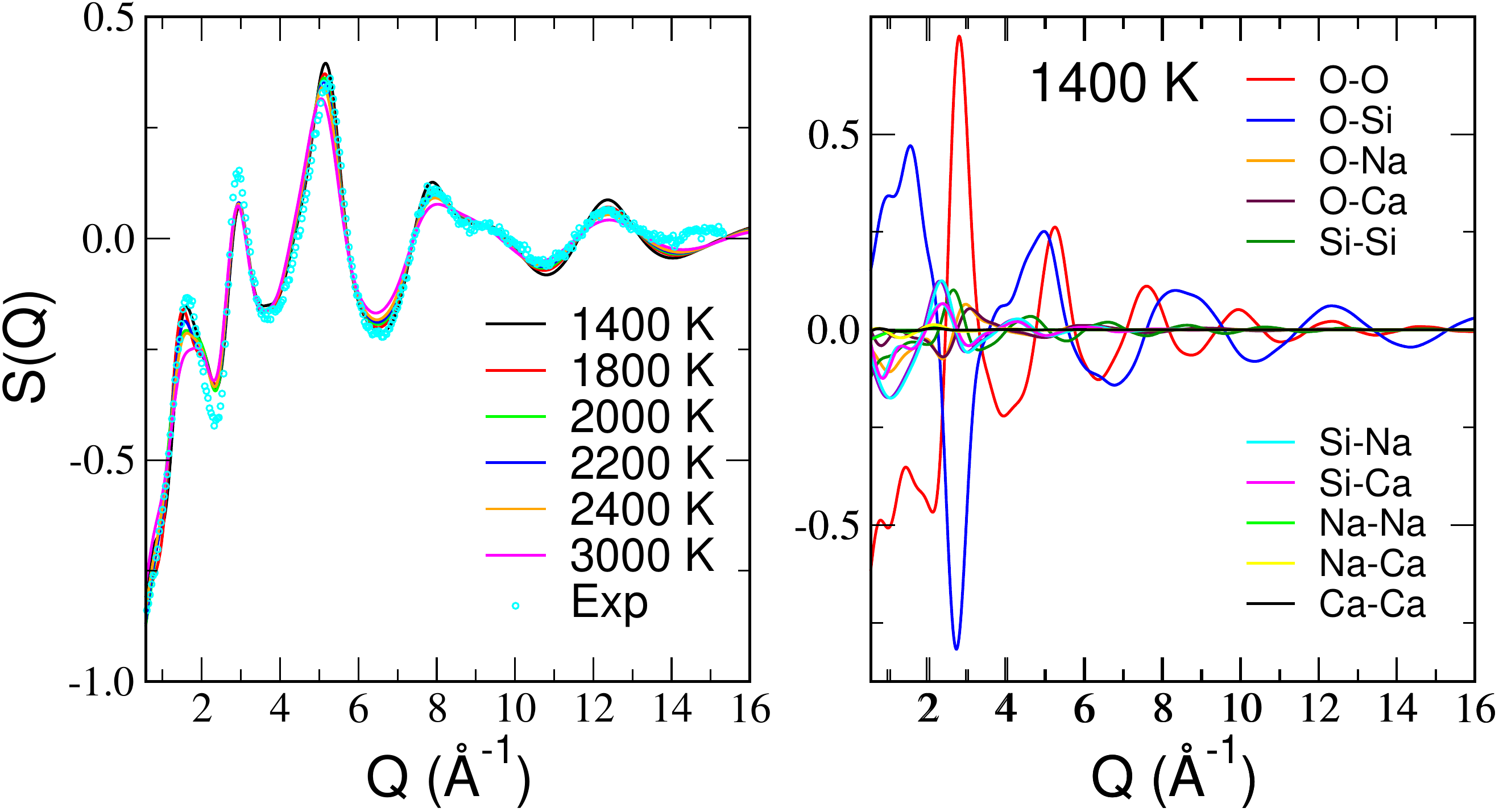}\\
\end{center}\caption{Left panel: calculated neutron weighted total structure factor as a function of temperature, and experimental structure factor at 1273 K from Ref.\citenum{cormier2011a}. Right panel: neutron weighted partial structure factors calculated at 1400~K.}\label{sq}
\end{figure}

In Figure \ref{sq} (left panel), the calculated neutron weighted structure factors, $S(Q)$, are plotted as a function of the temperature. They are obtained through the following equation: 
\begin{equation}
S(k)=1+\frac{1}{\mid\sum_\alpha c_\alpha w_\alpha(k)\mid^2}\sum_\alpha\sum_\beta c_\alpha c_\beta w_\alpha(k)w_\beta^*(k)[S_{\alpha\beta}(k)-1]
\end{equation}
\noindent where $k$ is the scattering vector magnitude, $c_\alpha$ is the atomic fraction (of chemical species $_\alpha$), $w_\alpha$ is the coherent neutron scattering length. $S_{\alpha\beta}$ are the partial structure factors, obtained from the partial RDFs (noted $g_{\alpha\beta}(r)$) by performing a Fourier transform:
\begin{equation}
S_{\alpha\beta}(k)=1+\frac{4\pi \rho}{k}\int_0^\infty r[g_{\alpha\beta}(r)-1]\sin(kr) {\rm d}r 
\end{equation} 
\noindent where $\rho$ is the atomic number density of the system. The total structure factor at 1400~K compares well with previous experiments carried out at 1273~K for a very similar system (75SiO$_2$-15Na$_2$O-10CaO).\cite{cormier2011a} Moreover, we can see that the high-$Q$ range, accounting for short-range interactions, is quite similar for all temperatures, while the low-$Q$ region is characterized by a peak at around 1.7~\AA$^{-1}$, which becomes broadened and less intense when the temperature increases. The decomposition in all the pair contributions is shown for a temperature of 1400~K in the right panel of Figure \ref{sq}. It is clear that the total structure factor is mainly dominated by the O-O and Si-O pairs, with Na-Na, Na-Ca and Ca-Ca correlations having only a small weighting factor. Note that all the partial $S(Q)$ involving the Al atom have not been reported as they provide a negligible contribution to the total signal, due to the very low concentration of Al in the system.

\subsection{Network structure}
\begin{table}[t]
\caption{Distribution (expressed in percentage) of the number of bridging (BO), non-bridging (NBO) and triple bonding oxygens and distribution of the Q$^n$ as a function of the temperature. See the text for more details.}
\label{bonbo}
\begin{center}
\begin{tabular}{lcccccccc}
\hline 
T(K) & BO & NBO & TBO & Q$^0$ & Q$^1$ & Q$^2$  & Q$^3$ & Q$^4$ \\
\hline 
1400 & 72.0 & 28.0 & 0    & 0 & 0    & 6.5 & 52.5 & 41.0 \\
1800 & 72.1 & 27.9 & 0 & 0 & 0.2 & 6.6 & 51.6 & 41.6 \\
2000 & 72.1 & 27.9 & 0 & 0 & 0.5 & 7.8 & 48.4 & 43.3 \\
2200 & 72.1 & 27.9 & 0 & 0 & 0.2 & 7.4 & 50.1 & 42.3 \\
2400 & 72.0 & 27.9 & 0.1 & 0 & 0.3 & 7.4 & 49.6 & 42.7 \\
3000 & 71.6 & 28.1 & 0.3 & 0 & 0.3 & 8.2 & 48.1 & 43.4 \\
\hline 
\end{tabular}
\end{center}
\end{table}
\indent In silicate glasses, the network structure is generally analyzed by splitting the oxygen population depending on their bonding. A bridging oxygen (BO) is defined as an oxygen atom connected to two network formers (Si/Al) in a sphere with radius corresponding to the first minimum of the Si-O/Al-O RDFs, while a non-bridging oxygen (NBO) is connected to only one Si/Al. Finally triple bonding oxygen (TBO) atoms are connected to 3 or more Si/Al atoms. The results of the analysis are summarized in Table \ref{bonbo} and show that the vast majority of oxygens are BO, with a ratio BO/NBO of about 2.6, independently of the temperature (the TBO percentage is almost negligible). \\
\indent Once BO, NBO and TBO have been identified, it is also possible to analyze the structure in terms of Q$^n$ distribution. Q is defined as a SiO$_4$ tetrahedra and $n$ is the number of oxygen atoms belonging to this tetrahedra that are BO. At the calculated BO/NBO ratio, one should expect that on average a Si tetrahedron has 1 NBO and 3BO, i.e. is a Q$^3$. Based on this, we should find in the Q$^n$ analysis a majority of Q$^3$, followed by a minority of Q$^4$ and Q$^2$, the latters having almost similar percentages. This is clearly not the case (see Table \ref{bonbo}), suggesting that the NBOs are not homogeneously distributed, but preferentially localized close to the network modifiers. For both BO/NBO and Q$^n$ distribution analysis, an overall agreement is obtained with the values previously determined for various modelling and simulation studies on systems with a similar composition.\cite{cormier2011a} \\
\subsection{Structure around the bridging/non-bridging oxygen atoms}

\begin{figure}[h!]
\begin{center}
\includegraphics[width=\columnwidth]{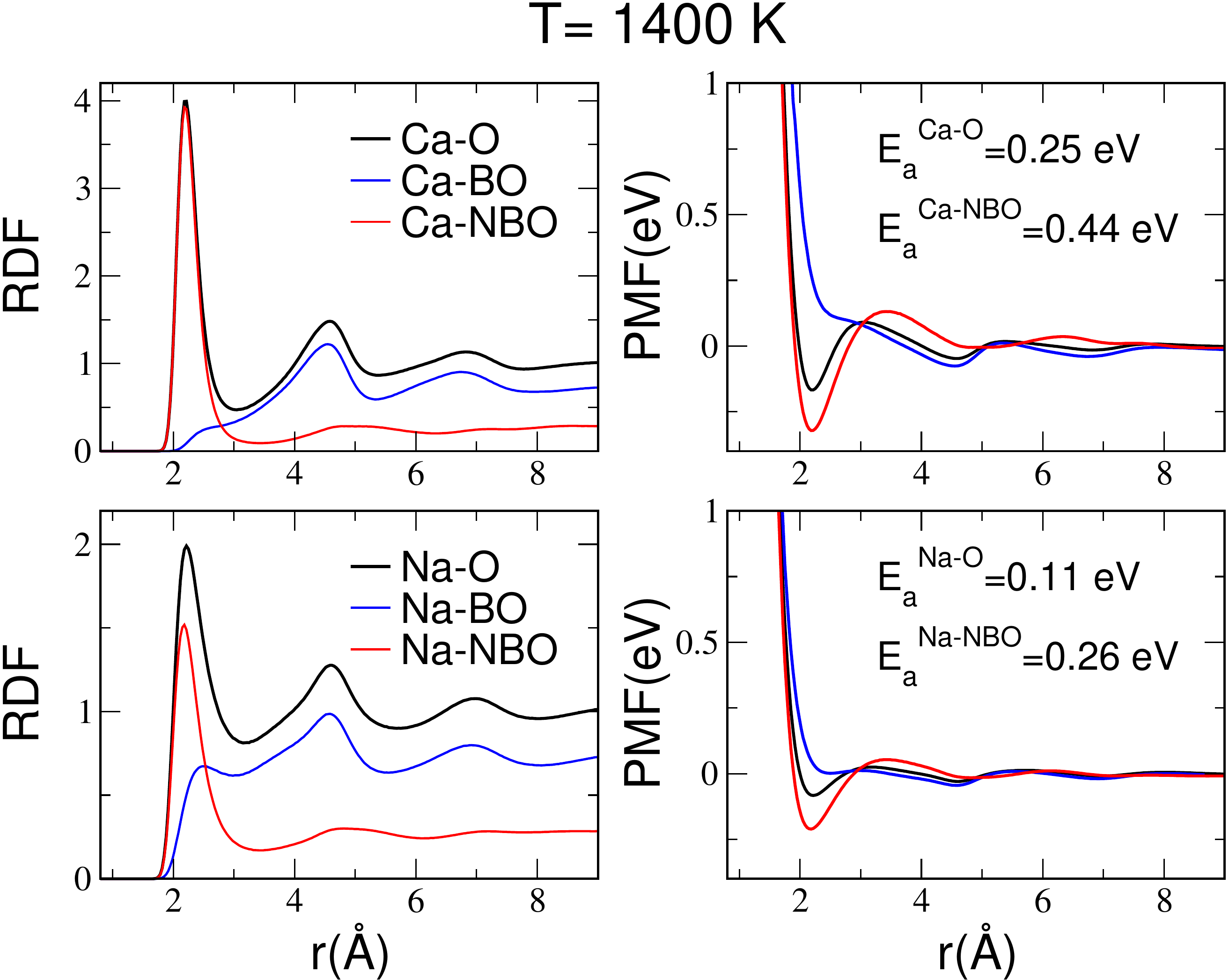}\\
\end{center}\caption{Left panels: Ca/Na-O, Ca/Na-BO and Ca/Na-NBO RDFs calculated at 1400~K. Right panels: corresponding potential of mean force (PMF, eV) extracted from the radial distribution functions.}\label{gdr_ca-na}
\end{figure}
\begin{figure}[t]
\begin{center}
\includegraphics[width=0.35\textwidth]{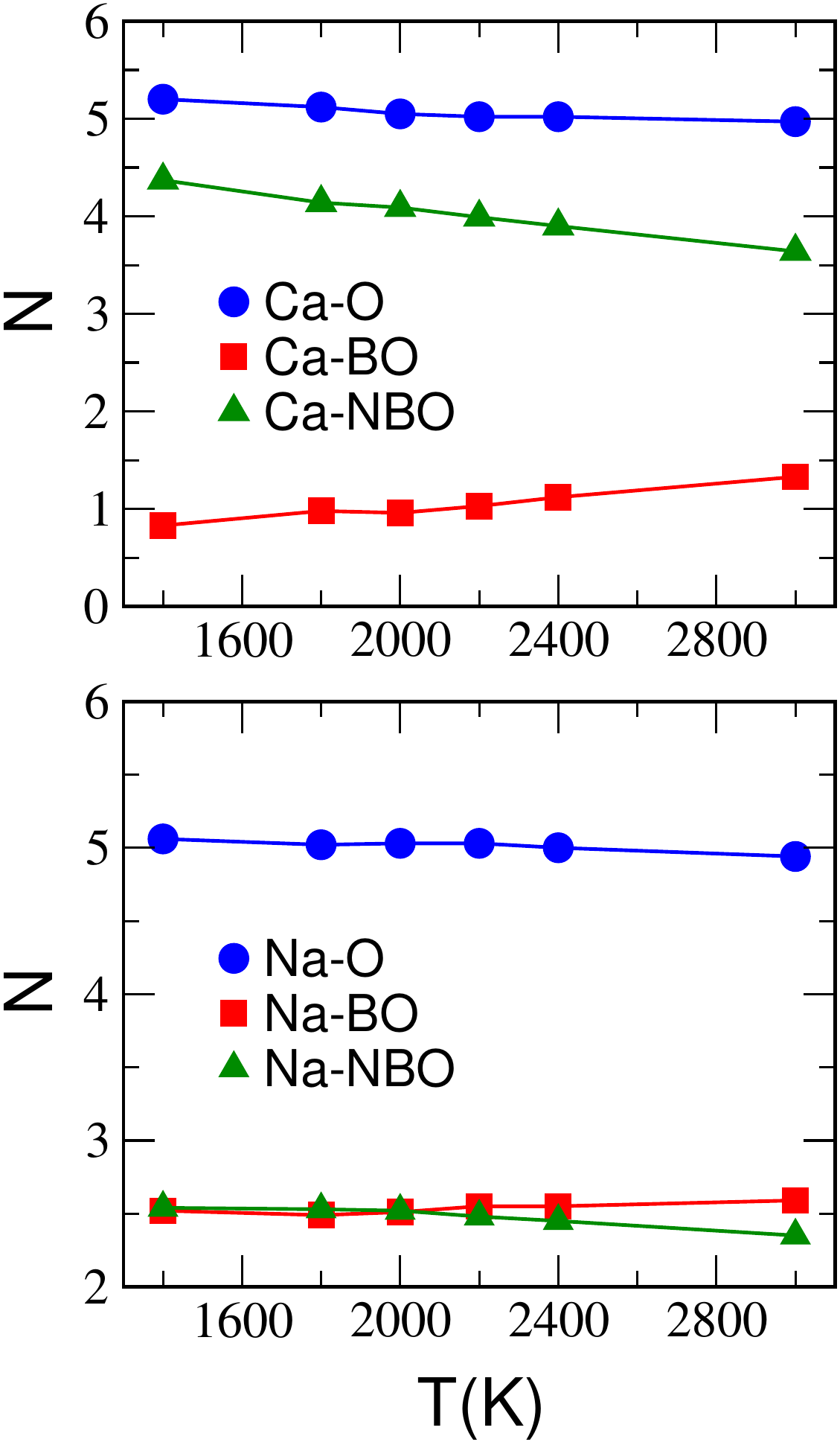}\\
\end{center}\caption{Average coordination number of oxygen, as well as its decomposition between BOs and NBOs, around Ca (top) and Na (bottom) as a function of the temperature. The cutoffs used are 3.09 and 3.23~\AA~for Ca and Na, respectively.}\label{ncoord_ca-na}
\end{figure}

The decomposition of the Ca-O and Na-O RDFs, shown in Figure~\ref{gdr_ca-na}
(left panels), clearly indicates that the first coordination shell of Ca is
largely formed by NBOs, with just a very small contribution from BOs. A
different situation is depicted for Na cations, where there is a more balanced
contribution of NBOs and BOs to the first coordination shell. Note that for
both cations, the Ca/Na-NBO RDF's first peak is found at shorter distances than
the Ca/Na-BO one. In particular, at 1400~K the Ca ions are on average
coordinated by 4.4 NBOs and 0.8 BOs, while there are 2.5 NBOs and 2.5 BOs
around Na cations (see Figure \ref{ncoord_ca-na}). The greater affinity of NBOs
for Ca than Na cations is preserved in the investigated temperature range, with
a very slight decrease in the number of NBOs from 1400~K to 3000~K accompanied
by a small increase in the number of BOs, in order to mantain the total average
coordination number almost constant. These observations agree with the
numerical study of Tilocca and de Leeuw,~\cite{tilocca2006} which also found a greater affinity of
NBOs for calcium than sodium ions, and with experimental NMR~\cite{jones2001}
or Raman~\cite{woelffel2015} results showing preferential arrangement of Ca in the vicinity of Q$^2$ rather Q$^3$ species. 

\subsection{Diffusion coefficients}
\begin{figure}[!h]
\begin{center}
\includegraphics[width=\columnwidth]{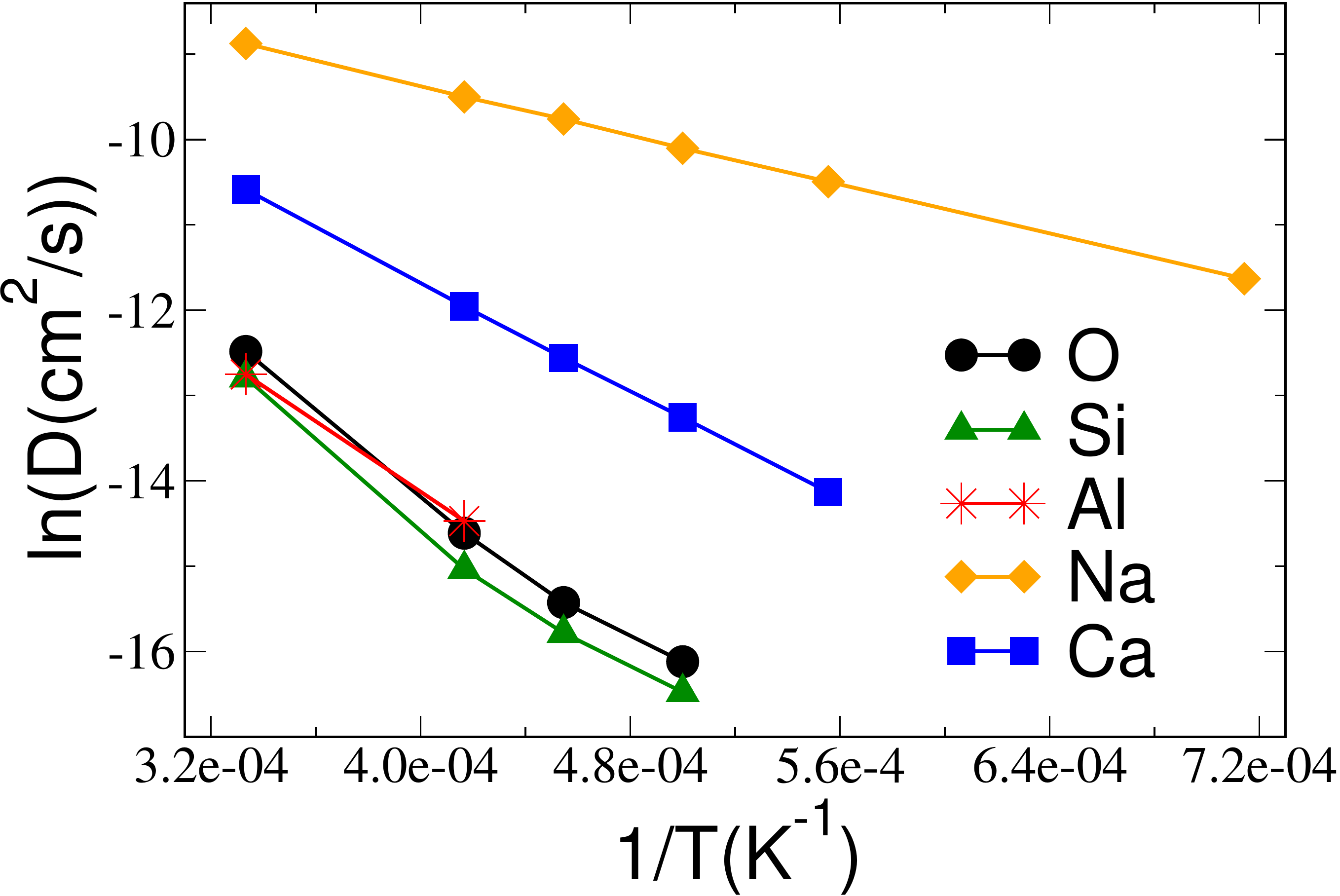}\\
\end{center}\caption{Diffusion coefficients plotted as a function of the inverse of temperature.}\label{diff}
\end{figure}

The self-diffusion coefficients have been calculated from the slope of the mean-squared displacements (MSD) versus time using the Einstein relation:
\begin{equation}
D_i = \lim_{t \rightarrow \infty} \dfrac{1}{6t}\langle\vert\delta\textbf{r}_a\left(t\right) \vert^2\rangle
\end{equation}
where $ \delta\textbf{r}_a $ is the displacement of a given ion of species $i$ in time $t$. \revadd{They are shown as an Arrhenius plot on Figure \ref{diff}.}\\
At high temperature ( $T$ $\geq$ 2000 K) all the species are in the diffusive
regime, with   Na and Ca cations showing a faster diffusion with respect to the
network formers (Si and O). It should be noted that the 
diffusion coefficient of aluminium is not easy to be determined due to the presence of 1 Al
atom only in the simulation box. At the lowest investigated temperatures, the
motion of the network modifiers starts to decouple from the one of the network
formers, and the Na and Ca cations diffuse into a silica matrix that is
basically frozen along all the simulation time. Finally, we can see that Na
ions always diffuse faster than the Ca ones, and both diffusion coefficients
follow a linear Arrhenius behaviour in all the investigated range of
temperature, i.e $D \propto e^{-E_{a}/RT}$, with an activation energy, $E_a$,
of 1.32 eV (127 kJ~mol$^{-1}$) and 0.62 eV (59.9
kJ~mol$^{-1}$) for Ca and Na, respectively. The activation energy
of Na is consistent with values found in the literature for similar
compositions.~\cite{pedone2008a} However, activation energies are significantly
smaller than the experimental values found closer to the glass transition.
~\cite{njiokep2008} Diffusion values of Si, Al and O are close to each other,
consistently with the literature observations that diffusivities of network
formers and oxygen are governed by the Eyring law relating diffusivity and
viscosity.\\

\indent The different diffusivities of Na and Ca can be linked to the  different structural organization adopted by these two cations. Indeed, the RDF and coordination number analysis at 1400~K (Figures \ref{gdr_ca-na} and \ref{ncoord_ca-na}) have shown that the difference in the first coordination shell of the two cations relies on the number of NBOs that they coordinate (4.4 for Ca $vs$ 2.5 for Na). From the RDFs it is also possible to calculate the potential of mean force (PMF) through the following expression:
\begin{equation}
	PMF = -K_BT  ln(g(r))
\end{equation}
where $ K_B $ is Boltzmann's constant and $T$ is the temperature of the system. Note that a maximum in RDF corresponds to a minimum in the PMF and {\it vice-versa}. The difference in energy between the first minimum and the first maximum in the PMF provides an estimate of the activation energy ($E_a$) needed to break one Ca/Na-O bond (see Figure \ref{gdr_ca-na}, right panels). The energy that a Na or Ca cation needs to escape from its first coordination shell can hence be estimated as:

\begin{equation}
E_{tot} = E_a \cdot N
\end{equation}
where $N$ is the average coordination number. When considering the total Ca-O and Na-O RDFs, we obtain E$_{tot}^{O}$ = 1.30 eV for Ca and E$_{tot}^{O}$ = 0.56 eV for Na. Thus, the Ca cation needs $\sim$ 2.5 times more energy to escape its coordination shell than Na. However, the difference in E$_{tot}^{O}$ may not arise from the total average coordination number, that is very similar between the two cations, but rather from their different NBO/BO ratio. This becomes evident when calculating the PMFs from the Ca/Na-BO and Ca/Na-NBO RDFs instead. As shown in Figure \ref{gdr_ca-na}, right panels, the PMF for Ca/Na-NBO exhibits a clear first minimum, which is deeper for Ca than Na, while the Ca/Na-BO PMF does not have any first minimum for both cations. In particular, E$_{tot}^{NBO}$ has been found equal to 1.92~eV and 0.66~eV for Ca and Na, respectively. This means that the energy cost to break the first solvation shell observed from both Na-O and Ca-O PMF solely comes from the Ca/Na-NBO bonds. It is therefore higher for Ca than Na because Ca is on average coordinated to more NBOs (4.4 vs 2.5 for Na) and with stronger bond (E$_a^{Ca-NBO}$ = 0.44 eV vs E$_a^{Na-NBO}$ = 0.26 eV).\\
\indent Interestingly, E$_{tot}^{O}$ obtained from PMF compares well with the activation energy previously determined from the diffusion coefficients, suggesting that escaping from the first oxygen coordination shell is the limiting step for the diffusion of Na and Ca cations. In the light of all these findings, the diffusion of the network modifiers is clearly ruled by the non-bridging oxygens coordinated to them, and thus Na diffuses faster than Ca because it is bonded to less NBOs and less strongly than Ca. This results also suggests that the diffusion mechanism is probably more interstitial-based (i.e an ion escapes its first coordination shell to an empty space in the network) than pair-based (i.e two cations exchange positions), the two mechanism being previously identified for the diffusion of network modifiers in silicate melts.\cite{pedone2008a,jund2001a,tilocca2010}\\

\section{Conclusion}\label{conclusion}
In this work a soda-lime-silica system in the liquid state (from 1400 to 3000
K) has been modelled by means of classical Molecular Dynamics simulations,
using an aspherical ion model that accounts for atomic polarization and
deformation effects. Overall, no structural important modifications occur going
from 1400 to 3000 K and the calculated neutron structure factor compares
well with previous experiments. In particular, we evaluated the structure in
terms of bridging (BO) and non-bridging oxygens (NBO), showing the greater
affinity of Ca$^{2+}$ ions for NBOs with respect to Na$^{+}$, which is
preserved in all the investigated temperature range. The different structural
organization adopted by the two cations has been further linked to their
different diffusivities. By computing the potential of mean force from the
radial distribution functions, we \revadd{find evidence} that the limiting step for the diffusion
of Ca$^{2+}$ and Na$^{+}$ ions is escaping from their first oxygen coordination
shell. This step requires more energy for Ca$^{2+}$ than Na$^{+}$, since Ca$^{2+}$ is on average coordinated to more NBOs and more strongly, thus the lower diffusivity of Ca$^{2+}$. \\
\indent Establishing at a molecular level the link between the structure and
diffusion properties of Ca$^{2+}$ and Na$^{+}$ ions is the first step for a
better understanding of glass melting and transformation processes. Some works
remain still to be done to further investigate if there are also multicomponent
diffusion effects~{\cite{liang2010, claireaux2016} that take place in such systems.

\begin{acknowledgements}

This work was supported by the French National Research Agency (project MAGI, Grant No. ANR-17-CE08-0019-04).	
	
\end{acknowledgements}

\section*{Data availability statement}
The data that support the findings of this study (input files for the simulations, raw data used for the various figures) are available \revadd{on a GitLab repository (https://gitlab.com/magi4) as well as on Zenodo (http://dx.doi.org/10.5281/zenodo.4266009)}.



%

\end{document}